\documentclass[letter,twocolumn]{jpsj2}
\usepackage[]{amsmath}
\usepackage[]{amssymb}
\usepackage[]{amsfonts}
\usepackage[]{latexsym}
\usepackage[]{float}
\usepackage[all]{xy}
\usepackage[]{endnotes}

\let\footnote=\endnote

\def\runauthor{Y. Hatsugai}

\title{%
Explicit Gauge Fixing for  Degenerate Multiplets:
\\
A Generic Setup for Topological Orders
}

\author{%
Yasuhiro  Hatsugai
\thanks{email:hatsugai@pothos.t.u-tokyo.ac.jp}
}

\inst{%
Department of Applied Physics, Univ. of Tokyo\\
7-3-1 Hongo, Bunkyo-ku, Tokyo 113-8656, JAPAN
}

\recdate{}

\abst{%
We supply  basic tools for the  study of the topological order
of a multiplet which is an eigenspace of a finite-dimensional normal 
operator with continuous parameters.
We allow intrinsic degeneracies within the multiplet where a well-known 
standard procedure does not work.
As an important example, we give novel expressions for a  spin Hall
conductance for unitary superconductors with equal spin pairing.
 Generic topological orders will be treated in this unified manner
 particularly with nontrivial topological degeneracies.
}

\kword{%
Chern numbers, degeneracy, topological orders
}

\newcommand{\mb}[1]{\mbox{\boldmath $#1$}}

\newcommand{\eas}[0]{\begin{eqnarray*}}
\newcommand{\eae}[0]{\end{eqnarray*}}
\newcommand{\les}[0]{\begin{equation}}
\newcommand{\lee}[0]{\end{equation}}
\newcommand{\leas}[0]{\begin{eqnarray}}
\newcommand{\leae}[0]{\end{eqnarray}}
\newcommand{\mchss}[4]
{
\left\{
\begin{array}{cc}
#1 & #2   \\
#3 & #4
\end{array}
\right.
}
\newcommand{\mat}[4]
{
\left(
\begin{array}{cc}
#1 & #2 \\
#3 & #4 
\end{array}
\right)
}

\newcommand{\mvec}[2]
{
\left(
\begin{array}{c}
#1  \\
#2  
\end{array}
\right)
}

\begin{document}

\maketitle


 It has been gradually clarified that
many  physically important phenomena 
have origins in their topological orders\cite{wen_topological1,geom-wl}.
Some of them   include (fractional and integer) 
quantum Hall effects\cite{kl-qhe,ando,aoki-ando,edge_l,tknn,streda,mkann,ntw,fqh_l,gmp},
Haldane spin chains\cite{haldane-spin}, 
solitons in polyacetylens\cite{ssh}, 
anisotropic
superconductors 
and superfluids \cite{volovik1,vol-book,smf,read-green,ym-yh-pairing,yh-sr}, 
chirality order in an itinerant magnetism\cite{tokura-chiral},
spin transport (spintronics) as a realistic application of 
Thouless pumping\cite{pump-th,spintronics},
polarizations in insulators\cite{ksv},
and the exotic electronic states of graphite\cite{sr-yh}.
Strong correlations between electrons cause exotic mean field 
states\cite{am,wwz}
 and effective quasiparticles such as composite fermions\cite{jain}
which can also be discussed 
in terms of the topological orders. 
In many cases, the topological order itself is hidden in  bulk systems 
but exhibits apparent physical consequences at the boundaries of the systems,
such as in edge states of
 the quantum Hall effects\cite{edge_l,edge_h,edge_yh},
local moments near impurities 
in the Haldane spin chains (Kennedy triplets)\cite{kennedy},
vortices and zero-bias conductance peaks 
in  anisotropic superconductivities
and boundary local moments in carbon nanotubes\cite{sr-yh}.

In many cases, nontrivial  topological  orders 
appear by restricting their physical space in a manner in which 
 a type of gauge 
structure naturally emerges\cite{dirac-gauge,berry, zee-wl}.
To characterize the quantum state of a specific system,
 one must explicitly determine gauge invariant quantities 
for the physical states. 
The (first) Chern number\cite{eguchi,zumino,mkann}
 is such a candidate and it has been used for several 
characterizations of topologically nontrivial states.
\cite{tknn,chern_yh,edge_yh,yh-qhe-review}

In this paper, we present a generic setup for 
the  discussion of the  topological order
explicitly,
particularly focusing on  gauge fixing. 
A standard procedure for fixing gauge was reported by
Kohmoto\cite{mkann}. This  is well known today.
However, the procedure does not work when degeneracy exists.
If degeneracy is accidental, that is, it exists at certain special
parameter values, it is negligible.
However,  
in several interesting situations such as in  unitary superconductors,
degeneracy is due to an intrinsic  symmetry, that is,
the standard procedure  cannot be applied for any values 
of the parameters (see below).
In such cases, the present generic gauge fixing procedure is essentially
important, particularly for numerical calculations of Chern numbers.
We extend the standard procedure to general situations which
allow  intrinsic degeneracies of eigenstates.
A typical situation where our method is
 crucial is the calculation of  spin Hall conductances
for numerically obtained BCS Hamiltonians, where
the order parameters are given numerically 
by minimizing the mean field free energy.
Then quasiparticle states are obtained by diagonalizing a
Bogoliuvov-de Gennes equation.
When the order is unitary,
it has an intrinsic degeneracy\cite{ueda-sg} which prevents   direct applications of 
the standard procedure to the calculation of the spin Hall conductance. 
Also, when a physical ground state has a nontrivial topological order
and it lives on a  genus $g(>0)$ Riemann surface,
a fundamental  topological degeneracy can occur with degeneracies
$q^g$ with some integer $q$\cite{wen_topological1}.
A typical situation is the fractional quantum Hall effect with the filling
factor $\nu=1/q$\cite{torus-q}.
In such a degenerate case, the present extension is indispensable.
Further generic expressions 
in the present paper can be  applicable to a wide range of
physically interesting situations.

\underline{Multiplet and Unitary Equivalence}:
Let us consider taking a
normal operator $\mb{L}$, 
($
\mb{L} ^\dagger \mb{L}   =
\mb{L}   \mb{L} ^\dagger 
$), in an  $N$-dimensional ($N<\infty$) linear space.
This implies that $\mb{L}$ is diagonalizable by 
a unitary matrix,  $\mb{\cal U}=(\psi_1,\cdots,\psi_N)$, as
$
\mb{L} \mb{\cal U}  = \mb{\cal U}  \mb{\cal E}
$,
$
\mb{\cal E }  = \text{diag}\, ( \epsilon_{1},\cdots,\epsilon_{N}  )
$.
Note that normal operators include
hermite, skew-hermite, unitary and skew-unitary operators.
(Correspondingly,
 the eigenvalues
$\epsilon_{i} $'s
 are real, pure imaginary, and on the unit circle
on the complex plane).
Also,
 we assume that
the operator $\mb{L}  $ is labeled by a set of continuous 
parameters as 
$\mb{L}(x) $, 
$x=(x_1,\cdots,x_{d=\text{dim}\, V})$,
 where $V$ is a $d$-dimensional parameter space\cite{fn1}. 
Various physical realizations of the operator
$ \mb{L} $ are 
(i)  momentum-dependent Hamiltonians 
in the quantum Hall effect\cite{tknn,mkann} and
an anisotropic superconductivity \cite{yh-sr,yh-chiral},
(ii) parameter-dependent Hamiltonians 
$\mb{L}(x)=  \mb{H} (x)$ 
in the discussion of  Thouless pumping\cite{pump-th}
and the Berry phase\cite{berry,zee-wl}, and  
(iii)  a time evolution operator, 
$\mb{L} (x)= T\exp(-(i/\hbar) \int dt H(t)) $.

Now construct an $M$-dimensional  multiplet 
(a linear space)  $W(x)$  with the parameters
which we considered
($\text{dim} W = M\le N$).
Take $M$ linearly independent orthonormalized bases $\psi_i$,
$i=1,\cdots,M$ for $W(x)$ ($M$ column vectors of dimension $N$)
and
 form an $N\times M$ matrix (the basis of the multiplet $W$) as 
\begin{alignat*}{1} 
\mb{\Psi} (x)=& (\psi_1,\psi_2,\cdots,\psi_M).
\end{alignat*} 
Supplementing the basis 
${\mb{\Psi}}_C $
of the orthogonal multiplet $W_C$  
($
W \oplus W_C = \mathbb{R}^N
$),
they form an orthonormalized complete basis for $\mathbb{R}^N $
as 
$
({\mb{\Psi}} , {\mb{\Psi}} _C) = \, \mb{\cal U} 
$.

Here, we must be cautious.
We require the multiplet $W(x)$ to be
uniquely specified by the parameter $x$.
However, this does not necessarily mean
that the basis ${\mb{\Psi}} (x)$ is uniquely specified by  $x$,
which gives us the freedom to change the basis. 
This leads to ambiguity in specifying the basis of $W(x)$.
That is, we may take a different basis, 
$ 
\mb{\Psi}'(x) =  \mb{\Psi} (x) \mb{\omega} (x)
$,
where $\mb{\omega}$ is an $M$-dimensional unitary matrix.
The ambiguity of the gauge freedom $\mb{\omega } $ can be  clarified,
for example,  when 
one tries to construct the multiplet from the $\psi_i$'s obtained numerically. 
This provides a gauge freedom for a connection we shall define.
This was first 
observed by Wilczek-Zee in a study on
non-Abelian Berry phases\cite{zee-wl}.

To specify the multiplet $W(x)$ 
uniquely using $x$,
the multiplet should  include all of the 
degenerate eigen spaces as 
\begin{alignat*}{1} 
 \epsilon_{i}(x) &\neq  \epsilon_{j}(x),
\\
i \in I^{in} = \{ 1,\cdots,M\}, &\
j \in I^{out}=\{ M+1,\cdots,N\}.
\tag{$*$}
\end{alignat*} 
We describe this condition the existence of a 
{\em generic energy gap}.
 The degeneracy within 
the multiplet is allowed, $\epsilon_{i}=\epsilon_{j},
i,j\in I^{in}$, 
which exists, for example, 
 in unitary superconductivity\cite{yh-chiral}
and spin degenerate cases.
The degeneracy between the multiplet $W$ we considered and
the supplementary $W_C$ is not allowed.

The parameter space and the multiplet are 
specified by the concrete  topological orders
we shall study.
Let us list some examples of $V$s with corresponding 
 multiplets $W$s:
 (1) $V$: the Brillouin zone,
(1-i) $W$: a collection of Landau level wave functions
in  the quantum Hall effects \cite{tknn},
(1-ii) $W$: quasi-particle states in 
anisotropic superconductors
 with and without equal spin pairing
\cite{yh-sr}, and
(2) $V$: a set of external parameters 
in the study of Thouless pumping\cite{pump-th} and 
the Berry phase\cite{berry,zee-wl},
 $W$: generically degenerate ground states, 
to be specific, for example,
 $V$: (2-i) a collection of fluxes passing through the systems,
Aharonov-Bohm fluxes and 
the strength of vortices in the type II superconductivity and  
(2-ii) parameters specifying the axes of  spins in magnetic systems.

\underline{Connection and  First Chern Number}:
Define a non-Abelian connection one-form $\mb{\cal A} $ 
which is an $M\times M$ matrix as 
$
\mb{\cal A} = \mb{\Psi} ^\dagger d \mb{\Psi} 
$.
Correspondingly, 
$
\mb{\cal A } ' = \mb{\Psi} ' d \mb{\Psi} ' 
= 
\mb{\omega } ^{-1}  \mb{\cal A} \mb{\omega }  + \mb{\omega } ^{-1} 
d \mb{\omega }
$.
This is a {\em gauge transformation} in our problem.
Also, define a field strength two-form
$
{\mb{\cal F}} = d {\mb{\cal A}} + {\mb{\cal A}} ^2
$
which is transformed  by gauge transformation as
$
{\mb{\cal F}} ' =
\mb{\omega } ^{-1}
{\mb{\cal F}} 
\mb{\omega } 
$.
Then ${\rm Tr \,} {\mb{\cal F}} $ is gauge invariant and 
an integral of ${\rm Tr \,} {\mb{\cal F}}  $ over the 
two-dimensional orientable compact surface $S$
($\partial S=0$)
in $V$
 gives
the first Chern number\cite{eguchi,zumino}
$
C_{S} = \frac {1}{2\pi i} \int_{S} {\rm Tr \,} {\mb{\cal F}} 
= \frac {1}{2\pi i} \int_{S} {\rm Tr \,} d {\mb{\cal A}} 
$.
For example, consider 
the three-dimensional Brillouin zone
$V=T^3(k_x,k_y,k_z)$ as 
 the full parameter space
and 
the  two-dimensional  Brillouin zone,
$S=T^2_{xy}(k_z )$, with the fixed third momentum $k_z$
for the surface $S$\cite{h3d,3dq,fn2}. 

A global connection over the full surface $S$
is not allowed to exist in a system 
with a nontrivial topological order. 
Then let us divide the integral region $S$ into several \underline{patches}
$S_R$ ($  S= \cup S_R, \quad R=0,1,2,\cdots $), 
where the connection ${\mb{\cal A}}_R $ is locally defined within $S_R$ as
${\mb{\cal A}}(x) = {\mb{\cal A}} _R(x)$\cite{eguchi,mkann}.
Furthermore,
 we assume each $S_R, R=1,2,\cdots,$ does not share any boundaries with
$
\partial S_0 = -\sum_{R\ge 1} \partial S_R
$.
When the connection ${\mb{\cal A}} _R$ is related 
to ${\mb{\cal A}} _0$ by the 
gauge transformation $\mb{\omega }_{0R} $ as 
$
{\mb{\cal A}} _R 
= \mb{\omega } _{0R} ^{-1} {\mb{\cal A}} _0 \mb{\omega } _{0R}
+\mb{\omega } _{0R} ^{-1} d  \mb{\omega } _{0R}
$,
the Chern number $C_{S}$ is written using the Stokes theorem as 
$C_{S}=
\frac {1}{2\pi }  \sum_{R\ge 1} \int_{\partial S_R}  {\rm Im\,}  {\rm Tr \,} 
 \mb{\omega } ^{-1} _{0R}  d \mb{\omega } _{0R}.
$

\underline{Explicit Gauge Fixing}:
Topological invariants are usually given 
by gauge-dependent quantities. 
To evaluate the expression, one must fix 
the gauge. 
Without fixing it,  we cannot have any well-defined derivative.
Now let us explicitly fix the gauge for the multiplet.
Although the basis ${\mb{\Psi}} $  has a gauge freedom,
a  projection  operator
 into the multiplet
 $ \mb{P}={\mb{\Psi}} {\mb{\Psi}} ^\dagger  $
is a gauge invariant.
Define an unnormalized basis,
${\mb{\Psi}}^U_\Phi $, from a generic 
 basis, ${\mb{\Phi}}$  (an $N\times M$ matrix ),
 as 
$
{\mb{\Psi}  }^U _\Phi
= \mb{P} {\mb{\Phi}} =
{\mb{\Psi}} \mb{\eta} _\Phi
$
($ \mb{\eta}_\Phi={\mb{\Psi}} ^\dagger \mb{\Phi} $).
The overlap matrix  of the basis
$
\mb{O}_\Phi =   {{\mb{\Psi}  }_\Phi^U} ^{\dagger} {\mb{\Psi}  }^U _\Phi
$
is generically  semipositive definite. 
Then, only if the determinant of the matrix $\mb{O}_\Phi $ 
 is nonzero,
 we can define a normalized wavefunction, 
$
{\overline{\mb{\Psi}  }} _\Phi
=  
 {\mb{\Psi}}_\Phi ^U \mb{o}_\Phi ^{-1} 
$,
where 
$
\mb{o} _\Phi \equiv
\mb{U} _\Phi^ {1/2}
\text{diag}({\sqrt{\lambda_1 }},\cdots,{\sqrt{\lambda_M }})
\mb{U} _\Phi 
$
with $ \mb{O}_\Phi =\mb{U} _\Phi 
\text{diag}(\lambda_1 ,\cdots,\lambda_M )
\mb{U} _\Phi$. 
This $\mb{o}_\Phi $ is hermite and positive definite.
Now we define the connection 
${\mb{\cal A}}_\Phi $ with the gauge fixing  by $\mb{\Phi} $ as 
${\mb{\cal A}} _\Phi =
 {\overline{\mb{\Psi}  }}_\Phi ^\dagger d {\overline{\mb{\Psi}  }} _\Phi
$.
This is well defined
 unless $\det \mb{\cal O}_\Phi=0  $.\cite{mkann,chern_yh}

Define regions
$S_R^\Phi, R=1,2,\cdots$,  as  (infinitesimally) small
neighborhoods  of 
zeros $x_R^\Phi$'s of  $\det \mb{\cal O}_\Phi(x) $ 
and  $S^\Phi_0$ as a rest of $S$ as
\begin{alignat*}{1}
S=\bigcup_{R\ge 0} S_R^\Phi,\quad \det \mb{\cal O}_\Phi (x)&
\mchss {\neq 0} { ^\forall x\in S^\Phi_0}
{ =0} { \text{at }^\exists x_R\in S^\Phi_R, R=1,2,\cdots .}
\end{alignat*} 
We use this gauge  for the region $S_0^\Phi$, and
for the region $S^\Phi_R$,  
we use a different gauge by $\tilde {\mb{\Phi}}, $ with
$\det \mb{O}_{\tilde \Phi} \neq 0$ everywhere in $S_R^\Phi$.
The transformation matrix  between 
${\overline{\mb{\Psi}}} _{\tilde\Phi} $ and 
${\overline{\mb{\Psi}}} _{\Phi} $ is obtained as 
$
\mb{\omega } = 
  \mb{o} _{\Phi} 
\mb{\eta} _{{\Phi}} ^{-1} 
 \mb{\eta} _{\tilde\Phi} 
\mb{o} _{\tilde\Phi} ^{-1} 
$.
Since 
${\mb{o}}_{\Phi} $ and  ${\mb{o}}_{\tilde\Phi} $ are strictly positive 
definite
at the boundaries  $\partial S^\Phi_0$, we have
$
{\rm Im\,} {\rm Tr \,}  \log \mb{\omega } 
=
-\, {\rm Im\,} {\rm Tr \,}
 \log \mb{\tilde\Phi}^\dagger  \mb{P}\mb{\Phi}
$. 
Finally, we obtain an expression for the first Chern number
with explicit gauge fixing as 
\begin{alignat*}{1} 
C_S =& - N_\Omega^T(S) =-\sum_{R\ge 1} n_\Omega^R(S_R^\Phi)\\
 n_\Omega^R(S_R^\Phi) =&   \frac {1}{2\pi} \oint _{\partial S^\Phi_R}
d'  \Omega ,
\quad\
\Omega = \Omega (\tilde {\mb{\Phi}} ,{\mb{\Phi}} )=
 {\rm Arg\,} 
\det \mb{\tilde\Phi}^\dagger  \mb{P}\mb{\Phi},
\end{alignat*} 
where $ N_\Omega^T(S) $
  is the total number  of signed vortices with the vorticity
 $ n_\Omega^R(S_R^\Phi) $
inside the region $S_R,\ R\ge 1$ 
( $\partial S^\Phi_0=-\cup_{R\ge 1} \partial S_R^\Phi$)\cite{fn3}.
Since 
$
\Omega=\text{Arg} \det \mb{\eta}_{\tilde \Phi} ^\dagger 
\det \mb{\eta} _\Phi  
$,
all the vortices of 
$\Omega(\tilde {\mb{\Phi}} ,{\mb{\Phi}} )$
 are given by zeros of $ \det \mb{\eta }_{\tilde \Phi}  $,
 ($x^{\tilde \Phi}$)
 and $ \det \mb{\eta }_\Phi  $,
 ($x^{\Phi}$).
The Chern number is obtained by summing up the vorticity only at
$x^\Phi$.
This form of  Chern number is not found in literature.

Since we assume that the two-dimensional surface $S$ is compact 
and 
$\Omega$ is regular
except at 
$x^\Phi_1,\cdots$ and 
$x^{\tilde\Phi}_1,\cdots$, 
a union of curves 
$(\cup_R S^\Phi_R )
\cup
( \cup_R S^{\tilde\Phi}_R)
$
 is contractible to a point
within a region  where 
$\Omega$ is well defined. 
This implies that
$
 N^T_{\Omega(\tilde{\mb{\Psi}},{\mb{\Psi}}  )}(S) =
  N^T_{\Omega({\mb{\Psi}},\tilde{\mb{\Psi}}  )}(S) 
$.
That is, the vector field $\Omega$ 
depends on  the gauge 
(choice of $\mb{\Phi} $ and
 $\tilde{\mb{\Phi}} $) but
the total vorticity $N_\Omega^T(S)$
  is a gauge invariant of the multiplet $W.$
None of the vortices has any direct physical meaning. 
Only the  total number of vortices $ N^T_\Omega(S)$
has a physical significance. 

The projection $\mb{P} $ is essential 
for carrying out
the present gauge fixing procedure. 
It  has also  an integral representation, 
$
\mb{P} =  \frac {1}{2\pi i} \oint_\Gamma G_T( z )
$,
where $\mb{G}_T= (\mb{I} _N - \mb{L} ) ^{-1} $ and 
the closed curve $\Gamma$  encloses
all of the eigenvalues $\epsilon_1,\cdots, \epsilon_{M}  $ inside, 
but not those $\epsilon_{M+1},\cdots, \epsilon_{N}  $
on the complex plane \cite{kato}.
From this form of projection, the stability condition of the generic gap $(*)$
for obtaining a well-defined multiplet is clear.
The first Chern number has 
an apparently gauge-independent form given by
$
{\rm Tr \,} {\mb{\cal F}}
= -{\rm Tr \,} d \mb{P}\mb{P}  d \mb{P} 
$ as well.\cite{avron-seiler-simon}
Also,
the Chern number for the multiplet is expressed as 
\begin{alignat*}{1} 
C_S =& 
\frac {1}{2\pi i } 
 \sum_{k\in I^{in}}
\int _S 
  \langle 
\, d\mb{L} ^\dagger  \, 
\{\mb{G}_C ( \epsilon_{k} ) \} ^\dagger 
\{\mb{G}_C ( \epsilon_{k} ) \}
\, d\mb{L}  
\rangle  _{\psi_k},n
\end{alignat*} 
where $\mb{G}_C= (\mb{I}-\mb{P} ) \mb{G}_T $.
This is equivalent to
a Kubo formula 
in the case of the quantum Hall effect\cite{tknn,avron-seiler-simon}.
This formula is particularly important since mathematical objects
such as Chern numbers have a direct relation with 
a physical quantity such as  a Hall conductance.
Surprisingly, this is  observable in a bulk system.

\underline{Sum Rules:} 
 Assume the multiplet  $W$ is a direct sum of orthogonal multiplets $W_1$ and $W_2$ as 
$
W = W_1 \oplus W_2
$,
which is expressed by  bases 
${\mb{\Psi}} _1$ and
${\mb{\Psi}} _2$ ( orthonormalized in each multiplet) as
$
{\mb{\Psi}} = 
 ({\mb{\Psi}} _1 ,{\mb{\Psi}}_2) 
$,
where
${\mb{\Psi}} _1 ^\dagger {\mb{\Psi}} _1  = \mb{I} _{M_1}$,
${\mb{\Psi}} _2 ^\dagger {\mb{\Psi}} _2  = \mb{I} _{M_2}$,
${\mb{\Psi}} _1 ^\dagger {\mb{\Psi}} _2 = \mb{O} _{M_1M_2}$,
and ${\mb{\Psi}} _2 ^\dagger {\mb{\Psi}} _1 = \mb{O} _{M_2M_1}$. 

The connection is given as 
$
{\mb{\cal A}} = 
\mvec
{{\mb{\Psi}} _1 ^\dagger }
{{\mb{\Psi}} _2 ^\dagger }
( d {\mb{\Psi}} _1, d{\mb{\Psi}} _2 ) 
$.
Thus, a trace of the connection is additive as
$
{\rm Tr \,} {\mb{\cal A}} =
 {\rm Tr \,}  {\mb{\cal A}} _1 + {\rm Tr \,} {\mb{\cal A}} _2
$, 
where 
${\mb{\cal A}}_1 = {\mb{\Psi}} _1 ^\dagger d {\mb{\Psi}} _1$ and 
${\mb{\cal A}}_2 = {\mb{\Psi}} _2 ^\dagger d {\mb{\Psi}}_2$.
From this simple observation in the connection level, {\em a sum rule} for Chern numbers 
is as follows:
$
C_S(W_1\oplus W_2) = 
C_S(W_1) +C_S(W_2)
$.
The sum rule in 
the field strength level was previously discussed\cite{avron-seiler}.
A simple consequence of the present sum rule  is the total sum rule, that is,
the Chern number of the total multiplet $W_T$ always vanishes,
\cite{yh-chiral}
$
 \sum_i C_S( W_i) 
=
C_S(\oplus_i W_i)
=
0
$,
since
$
\mb{P}_{\oplus_i W_i } =  \mb{I}_N
$.


\underline{One-Dimensional Example ($\text{dim } W=1$)}:
When the multiplet is one-dimensional, such
as
$
{\mb{\Psi}} = \psi
$ 
and 
$
(\mb{P} )_{ij} =  \psi_i\psi_j ^*
$,
we have
$
\tilde{\mb{\Phi}} ^\dagger  \mb{P} \mb{\Phi}
  =  (\mb{P}  )_{N1} = \psi_N^*\psi_1
$
by taking $\mb{\Phi}$ and $\tilde {\mb{ \Phi}}$ as 
$
^t {\mb{\Phi}} = 
({1},{0},{\cdots}{0})
$ and 
$^t{\tilde{\mb{\Phi}}} = 
({0},{\cdots},{0},{1})$.
Then the Chern number is given as  
\begin{alignat*}{1} 
C_S =& 
-\frac {1}{2\pi}
 \oint_{\partial S^\Phi_0}d\,  {\rm Im\,} \log ( {\psi_1}/{\psi_N}),
 \quad\
S^\Phi_0 = S\setminus\bigcup\nolimits_{R\ge 1} S^\Phi_R,
\end{alignat*} 
where $S^\Phi_R$ includes a single zero of 
$\det \mb{\cal O}_\Phi=|\psi_1|^2 $.
This is a well-known classic expression\cite{mkann}.

\underline{Multiplet of Several Landau Levels}:
When one considers two-dimensional electrons on a lattice with
the flux $\phi$ per plaquette, one-particle states are given by $q$ bands
when $\phi=p/q$ with the mutually prime $p$ and $q$. Furthermore, the spectrum is
given by the famous  Hofstadter's butterfly. When the fermi energy
$E_F$
is in the 
$j$-th energy gap from below, the Hall conductance $\sigma$ is 
given by the sum of the Chern numbers of the $j$ bands\cite{tknn}. 
In this case, take a multiplet from a filled fermi sea ($W=\text{FS}$) and 
construct the basis of the multiplet from the $j$ Bloch states
$\psi_j( \mb{k}) $ below 
$E_F$ as ${\mb{\Psi}} =(\psi_1,\cdots,\psi_j) $,
$M=j$, then the Chern number $C_{T^2}(\text{FS})$ naturally gives
the Hall conductance $\sigma _{xy}$ 
which is the  sum of the Chern numbers of 
the filled bands\cite{tknn,avron-seiler,chern_yh,yh-qhe-review}.

\underline{Dirac Monopole}:
When the dimension of the total Hilbert space $N$ is 2,
only the nontrivial multiplet is  one-dimensional $M=1$.
Then take  an hermite Hamiltonian
$
\mb{H}( x) = \mb{R}(x) \cdot \mb{\sigma } 
$
for the normal operator $\mb{L}$ 
where  $\mb{\sigma } $'s are Pauli matrices and $\mb{R} (x)$ is 
a real three-dimensional vector
( $(R,\theta,\phi)$ is a polar coordinate of $\mb{R}$).
As an  example, consider the multiplet ${\mb{\Psi}}_- $ with the energy 
$- R $ as 
\cite{berry}
$
^t{\mb{\Psi}}_-(x) = 
(
{-\sin \frac {\theta(x)}{2} },
{e^{i\phi(x)}\cos \frac {\theta(x)}{2} }
)
$.
The projection is given as 
$
\mb{P} _- = 
{\mb{\Psi}}_- {\mb{\Psi}}_- ^\dagger 
= 
\mat
{\sin^2 \frac {\theta}{2} }
{-e^{-i\phi}\sin \frac {\theta}{2} \cos \frac {\theta}{2} }
{-e^{i\phi}\sin \frac {\theta}{2} \cos \frac {\theta}{2} }
{\cos^2 \frac {\theta}{2} }
$.
Using  a gauge by $\mb{\Phi} $ and $\mb{\tilde\Phi} $
as
$
^t\mb{\Phi}  = 
(
{ \cos \frac \chi 2 },
{e^{i\xi}\sin \frac \chi 2 })
$ and
$
^t\mb{\tilde\Phi}  = 
(
{\cos \frac {\tilde\chi}2 },
{e^{i\tilde\xi}\sin \frac {\tilde\chi} 2 })
$,
$\det \mb{\cal O}_\Phi=0 $ and 
$\det\mb{\cal O}_{\tilde{\Phi}}=0 $ give
$
(\theta(x),\phi(x)) = 
{(\chi,\xi)}
$ 
and 
${(\tilde\chi,\tilde\xi)}$,
 respectively.
This clearly shows that the positions of the vortices defined 
by the vector field
$\Omega=\text{Arg}
(
- \sin \frac {\theta}{2}  \cos \frac {\tilde\chi}{2} 
+
e^{+i(\phi-\tilde\xi)} \cos \frac {\theta}{2}  \sin \frac {\tilde\chi}{2} 
)
(
- \sin \frac {\theta}{2}  \cos \frac {\chi}{2} 
+
e^{-i(\phi-\xi)} \cos \frac {\theta}{2}  \sin \frac {\chi}{2} 
)
$
are gauge-dependent
 and do not have any direct physical meaning.
One can chose the positions of the vortices as one wishes.


\underline{Unitary Superconductors}:
Let us first consider the simplest case,
that is, the unit cell includes only one site.
Then the  Bogoliuvov-de Gennes equation for generic superconductivity
is given  in a momentum space by a $4\times 4$ secular equation, 
$
\mb{H} \psi = E\psi
$,
$
\mb{H}  = 
\mat
{\epsilon \mb{I}_2} {\mb{\Delta}}  
{\mb{\Delta ^\dagger }}  {-\epsilon\mb{I}_2}
$.
As for the unitary order, the order parameter matrix $\mb{\Delta_{} } $
 is written as 
$
\mb{\Delta} \equiv 
|\Delta| \mb{\Delta}_{0},\quad | \Delta_{} |\ge 0$,
$
\mb{\Delta }\mb{\Delta } ^\dagger = |\Delta| ^2 \mb{I}_2
$,
where $ \mb{\Delta}_{0} $ is a 2$\times 2$ unitary matrix.
Then the eigenstates (quasiparticle)
 are doubly degenerate as
$
^t \psi_-(\mb{w} ) = 
(
{-\sin \frac {\theta}{2} \, \mb{w} },
{\cos \frac {\theta}{2} \, \mb{\Delta }_0 ^{^\dagger }  \mb{w} },
)
$, 
for example, for  the $E = - R$ state
where 
$\mb{w} $ is a normalized arbitrary two-component vector,
 $\mb{w}^\dagger \mb{w}=1  $,
$R=\sqrt{|\Delta|^2+ \epsilon^2}$,
$\epsilon = R \cos \theta $ and 
$|\Delta| = R \sin \theta $.
Now let us construct a multiplet for the degenerate $E=- R$ quasiparticle bands 
as 
$
{\mb{\Psi}}_- =
 ( \psi_-(\mb{w}_1 ),\psi_-(\mb{w}_2 ))
$, 
where $\mb{w}_1 $ and $\mb{w}_2 $ form an 
arbitrary two-dimensional orthonormalized complete set:
$
\mb{w} _i ^\dagger \mb{w} _j = \delta_{ij} 
$,
$
\mb{w} _1  \mb{w} _1 ^\dagger 
+
\mb{w} _2  \mb{w} _2 ^\dagger 
 = \mb{I} _2
$.
Then the projections $\mb{P}_- $  are given in a gauge invariant form as
$ 
\mb{P} _- 
=
\mat 
{ \mb{I}_2\sin ^2 \frac \theta 2}
{-\mb{\Delta }_0 \sin \frac {\theta}{2}\cos \frac {\theta}{2}   }
{-\mb{\Delta }_0 ^\dagger  \sin \frac {\theta}{2}\cos \frac {\theta}{2}    }
{ \mb{I}_2\cos ^2 \frac {\theta}{2}}
$. 
Now let us fix the gauge by choosing 
$^t\mb{\Phi} = ^t({\mb{0}_2  },{\mb{I}_2})$ and 
$^t\tilde {\mb{\Phi}} =^t({\mb{I}_2},{\mb{0}_2  })$.
Then we have
$
\det \mb{\cal O} _{-,\Phi} = 
{ \cos^4 \frac {\theta}{2} }
$,
$\Omega= \text{Arg} \det \mb{\Delta}_{0} =\text{Arg} \det \mb{\Delta}
$.
Since the overlap determinant
 $\det \mb{O} _{-,\Phi} =0$ vanishes at $\theta=\pi$ 
 for the multiplet $W_-$,
the Chern number is given by
\begin{alignat*}{1} 
C_{ S}^- =& -\frac {1}{2\pi} \sum_p\oint _{p} 
d' \text{Arg} \det \mb{\Delta},
\end{alignat*} 
where  $p$'s are points on the surface $S$ which are specified by $\theta=\pi$,
that is,
$|\Delta|(p)=0$ and $\epsilon(p)=-R(p)=-E(p)$.
This is a novel expression for the 
generic spin Hall conductance for the unitary superconductors
with equal spin pairing.
In the previous work\cite{yh-chiral}, 
 the Chern number was given as the sum of two integers  using an 
eigenvalue equation for the unitary matrix $\mb{\Delta}_0 $.
Here, we give a direct expression only using
 the order parameter matrix $\mb{\Delta}$.
Furthermore, if one parameterizes the unitary 
 $2 \times 2$ matrix $\mb{\Delta}_{0}$ as 
$
\mb{\Delta}_0  = e^{i\Theta } e^{i \hat n\cdot \vec  \sigma },\ |\hat n| =1
$, we have
$
C_{ S}^- = -\frac {1}{\pi} \sum_p\oint   _{p} d \Theta
$.
The present method is crucially important and efficient when 
the order parameter is given by  numerically solving a BCS self-consistent equation
with a large unit cell.
Even in this generic situation,
 to evaluate the spin Hall conductance,
we must determine arbitrary orthonormalized (degenerate) eigenstates 
by diagonalizing the mean field Hamiltonian
(Bogoliuvov-de Gennes equation).

Finally, we mention  the higher order Chern numbers
 $C_n(S^n)=N_n\int_{S^n}{\rm Tr \,}  {\mb{\cal F}} ^n ,n=2,3,\cdots$,
where $N_n$ is a normalization constant
and  $S^n=S\times\cdots S$, ($n$-times).
In principle, 
they can also explicitly be evaluated  using the present gauge fixing procedure
as a sum of integrals over the $(2n-1)$-dimensional spheres 
$S_R^\Phi$ enclosing
$(2n-2)$-dimensional regions $P_R^\Phi$ which are defined by the
zero of $\det \mb{\cal O}_\Phi $ in $S^n$.
 They should  also help in the characterization of 
the topological order 
in  complex situations.

We thank M. Kohmoto for useful discussions.
Part of the  work by Y.H. was supported by a Grant-in-Aid from the
Japanese Ministry of Education, Culture, Sport, Science and Technology, and 
the JFE 21st Century Foundation.


\begin{thebibliography}{43}
\bibitem{fn2}
The surface $S$ can be a genus $g$ Riemann surface $\Sigma_g$.
The simplest and the most common case is the torus $T^2=\Sigma_1$ 
which appears as a  two-dimensional  Brillouin zone and 
a couple of fluxes in the multiflux AB effects.
\bibitem{fn3}
We have to note that integral theorems 
 cannot be
applied here since $\log \det \mb{\omega }  $ and $ {\rm Arg\,} 
\det \mb{\tilde\Phi}^\dagger  \mb{P}\mb{\Phi}$ are not  global 
single valued functions at the boundaries $\partial S_0^\Phi$.
We use $d'$ to indicate it.
\end{thebibliography}
\end{document}